\documentclass
{mn2e}
\usepackage{epsfig}
\usepackage{amsmath, amssymb,bm}

\def \msun{\rm M_{\odot}}

\begin{document}
\title[AGN Flickering and Chaotic Accretion]{AGN Flickering and Chaotic Accretion}

\author[Andrew King \& Chris Nixon] 
{
\parbox{5in}{Andrew King$^{1}$ \& Chris Nixon$^2$}
\vspace{0.1in} \\ $^1$Department of Physics \& Astronomy, University
of Leicester, Leicester LE1 7RH UK \\ $^2$JILA, University of Colorado \& NIST, Boulder CO 80309-0440, USA}

\maketitle

\begin{abstract}
Observational arguments suggest that the growth phases of the supermassive black holes in active galactic nuclei have a characteristic timescale $\sim 10^5$~yr. We show that this is the timescale expected in the chaotic accretion picture of black hole feeding, because of the effect of self--gravity in limiting the mass of any accretion disc feeding event.
\end{abstract}

\begin{keywords}
{galaxies: active: galaxies: Seyfert:  quasars: general: quasars: supermassive black holes: black hole physics: X--rays: galaxies}
\end{keywords}

\footnotetext[1]{E-mail: ark@astro.le.ac.uk}

\section{Introduction}
\label{intro}
In a recent paper, Schawinski et al. (2015; hereafter SKBS) show that nearby
active galactic nuclei (AGN) undergo cycles (`flickering') with a characteristic timescale. The cycle starts as X--ray emission turns on. Until enough time elapses for X--ray and UV photoionization to produce the optical appearance of an AGN, the system is in the  `optically elusive AGN' or `X--ray Bright Optically Inactive Galaxy' (XBONG) phase. Once photoionization 
is achieved, the usual AGN phase follows, with both X--ray and optical signatures present. Most 
of the growth of the supermassive black hole (SMBH) occurs in this phase. This ends as the X--rays turn off, and a light echo phase (as seen in `Hanny's Voorwerp') may appear after it. SKBS show that the expected duration of photoionization, together with the observed fraction of optically elusive AGN, give an estimate of the typical lifetime of the AGN phase of a galaxy as 
\begin{equation}
t_{\rm AGN} \sim 10^5~{\rm yr}. 
\label{tagn}
\end{equation}
This number appears to be effectively independent of the galaxy sample used (SKBS). Note that this argument says nothing about the recurrence timescale of the growth events, only their duration.

In this paper we show that the picture of chaotic accretion on to SMBH introduced by Sanders (1981) and developed by King \& Pringle (2006) gives a simple explanation for both the characteristic growth timescale (\ref{tagn}) and its similarity for all types of nearby AGN. 

\section{Chaotic Accretion}
\label{chaos}
Sanders (1981) first suggested a picture of accretion on to a supermassive black hole
which occurs via a sequence of randomly oriented accretion discs. Several authors have since revived this idea (e.g. Moderski, Sikora \& Lasota 1998, in discussing the radio properties of quasars). King \& Pringle (2006)  pointed out that it readily accounts for the lack of any correlation between the directions of AGN jets (and so presumably SMBH spin axes) and the geometry of the host galaxy, and second that it allows the spin rate of the SMBH, and so its accretion efficiency, to remain low. This second point in turn means that the Eddington limit corresponds to a relatively high mass accretion rate, allowing rapid growth of the SMBH mass to the values $\sim 10^9 - 10^{10}\msun$ observed at redshifts $z\sim 6$ (e.g. Barth et al. 2003; Willott, McLure \& Jarvis 2003; Wang et al., 2015) even from a stellar initial mass. 

King \& Pringle (2006) showed that the necessary condition for low SMBH spin is that most accretion disc episodes have total angular momentum $J_d \la J_h$, where $J_h$ is the hole's spin angular  momentum. This condition means that the disc and SMBH spin end up counteraligned in about one--half of all cases. The greater lever--arm of the innermost stable circular orbit (ISCO) in retrograde cases means that the net effect of randomly--oriented accretion events is to spin the hole down. In calculating the predicted spin and mass evolution of SMBH, King, Pringle \& Hofmann (2008) showed that
 the requirement $J_d \la J_h$ is in general easily satisfied, since the total mass of a 
disc episode is limited by the self--gravity constraint (Pringle, 1981; Frank, King \& 
Raine, 2002)
\begin{equation}
M_d \sim {H\over R}M.
\label{sg}
\end{equation}
Here $H/R$ is the disc aspect ratio, and $M$ the SMBH mass. This constraint means that discs become self--gravitating outside some radius $R_{\rm sg} \sim 0.01$~pc (Collin-Souffrin \& Dumont 1990; Shlosman et al. 1990; Hur\'e et al. 1994; King \& Pringle 2007). Cooling times in these regions are so short that this
is likely to result in star formation rather than accretion (Shlosman \& Begelman 1989; Collin \& Zahn 1999). 

The importance of the self--gravity constraint goes wider.  As discussed by King \& Pringle (2007), disc self--gravity is a fundamental barrier to accretion: no disc event forming outside $R_{\rm sg}$  -- i.e. with initial specific angular momentum $\ga (GMR_{\rm sg})^{1/2}$ -- can easily contribute to central accretion. We expect that most of the gas initially at radii $R >
R_{\rm sg}$ either forms stars, or is expelled by those stars which do
form on a rapid (almost dynamical) timescale.  Gas
initially at radii $R < R_{\rm sg}$ must form a standard accretion disc, 
gradually draining on to the black hole and powering the AGN. 
King \& Pringle (2007)
show that for local, low--luminosity AGN, fuelling by well--separated
episodes of this type explains observational features such as the luminosity
function for moderate--mass black holes, and the presence
and location of a ring of young stars observed about the Galactic
Centre.

We conclude that all SMBH feeding must largely come from discs with initial specific angular momentum small enough for them to form at radii $R \la R_{\rm sg}$. As a result, disc events cannot last longer than the time for accretion to drain the mass from within $R_{\rm sg}$. So the maximum duration of a disc event is 
\begin{equation}
t_{\rm sg} \sim {M_d\over \dot M} \sim {HM\over R\dot M}
\label{var}
\end{equation}
where $\dot M$ is the average accretion rate.

In AGN discs, the aspect ratio is always small, and almost independent of parameters, i.e.
\begin{equation}
\frac{H}{R} = 1.94 \times 10^{-3}\alpha_{0.03}^{-1/10}\eta_{0.1}^{-1/5} 
\dot m^{1/5} M_8^{-1/10} r^{1/20},
\label{thickness}
\end{equation}
(cf King, Pringle \& Hofmann 2008).
Here $\alpha = 0.1\alpha_{0.1}, \eta = 0.1\eta_{0.1},  \dot m = \dot M/\dot M_{\rm Edd}$ are the standard viscosity parameter, the accretion efficiency and Eddington accretion ratio respectively, and $M_8 = M/10^8\msun, r = R/R_g$, with $R_g$ the gravitational radius $GM/c^2$.\footnote{Larger values of $H/R$ can be produced if the disc is assumed to be bathed in AGN emission reradiated by a very close--in warm absorber of similar size $\la 10^4$ Schwarzschild radii, cf Loska, Czerny \& Szczerba, 2004. However recent observations by Tombesi et al., 2013 suggest that the likely distance of such components is a factor $\sim 10^4$ greater, making the reradiation effect negligible.}
Since $H/R$ is insensitive to any of these parameters we take $H/R \simeq 2\times 10^{-3}$, and so (\ref{var}) becomes 
\begin{equation}
t_{\rm sg} \sim 10^{-3}t_{\rm Sal}\dot m^{-1} \sim 10^5\dot m^{-1}~{\rm yr},
\label{var2}
\end{equation}
where $t_{\rm Sal} = M/\dot M_{\rm Edd} \simeq 4\times 10^7\eta_{0.1}$\,yr
is the Salpeter timescale. A full calculation (cf King \& Pringle, 2007) gives 
\begin{equation}
 t_{\rm sg} \sim 1.7\times 10^5\alpha_{0.1}^{-2/27}\eta_{0.1}^{22/27}\dot m^{-22/27}M_8^{-4/27}~{\rm yr}
 \label{var3}
 \end{equation}
King \& Pringle (2007) show that an AGN  luminosity function close to the observed one (Heckman et al., 2004) follows if the initial value of $\dot m$ is $\sim 1$, fixing the luminosity evolution of each disc event as
\begin{equation}
L \simeq L_{\rm Edd}[1 + (t/t_{\rm var})]^{-19/16}
\label{lum}
\end{equation}
where $L_{\rm Edd}$ is the Eddington luminosity and $t_{\rm var}$ is $t_{\rm sg}$ with $\dot m =1$, i.e.
\begin{equation}
 t_{\rm var} \sim 1.7\times 10^5\alpha_{0.1}^{-2/27}\eta_{0.1}^{22/27}
 M_8^{-4/27}~{\rm yr}
\end{equation}
So chaotic accretion predicts a mass--growth timescale $t_{\rm var}$ of AGN variation very close to what SKBS deduce from observations. This result is almost independent of AGN parameters. AGN may of course vary in other ways (e.g. through the thermal--viscous disc instability, cf Burderi et al., 1998) but only if these have shorter timescales than the mass--supply time $t_{\rm var}$. Such mechanisms cannot have a major effect on SMBH mass growth, since the self--gravity constraint implies that much of this is at near--Eddington rates (cf eqn \ref{lum} above).

\section{Discussion}

The simplicity of the derivation of the timescale $t_{\rm var}$ is striking. It depends only on the fact that the disc aspect ratio $H/R$ is always $\sim 10^{-3}$ for AGN accretion discs, and so is an immediate and inescapable consequence of chaotic accretion. The very tight limit  $R_{\rm sg} \la 10^{-2}$~pc 
on the formation radius of AGN discs, as derived by King \& Pringle (2007), 
is independently supported by the variation timescale that SKBS deduce from observation. It implies that gas accreting on to SMBH must have been `aimed' very precisely towards them, and is significantly tighter than the limit $R_{\rm circ} \la 1$~pc needed for the viscous timescale to be less than the Hubble time. A potential explanation for these stringent limits may be that SMBH feedback itself tends to provoke very radially--directed accretion (Dehnen \& King, 2013). Whatever the mechanism driving accretion, the emerging picture of SMBH feeding appears to suggest that it is likely to be chaotic on small scales of length, time and mass, rather than orderly.

\section*{Acknowledgments}
 Research in theoretical astrophysics at Leicester is supported by an STFC Consolidated Grant. CN was supported for this work by NASA through the Einstein Fellowship Program, grant PF2-130098.



\begin{thebibliography}{}


\bibitem[\protect\citeauthoryear{Barth et al.}{2003}]{2003ApJ...594L..95B} 
{Barth A.~J., Martini P., Nelson C.~H., Ho L.~C., 2003, ApJ, 594, L95} 

\bibitem[\protect\citeauthoryear{Burderi, King, 
\& Szuszkiewicz}{1998}]{1998ApJ...509...85B} Burderi L., King A.~R., Szuszkiewicz E., 1998, ApJ, 509, 85 


\bibitem[\protect\citeauthoryear{Collin 
\& Zahn}{1999}]{1999A&A...344..433C} Collin S., Zahn J.-P., 1999, A\&A, 344, 433 


\bibitem[\protect\citeauthoryear{Collin-Souffrin 
\& Dumont}{1990}]{1990A&A...229..292C} Collin-Souffrin S., Dumont A.~M., 1990, A\&A, 229, 292 


\bibitem[\protect\citeauthoryear{Dehnen 
\& King}{2013}]{2013ApJ...777L..28D} Dehnen W., King A., 2013, ApJ, 777, L28 



\bibitem[\protect\citeauthoryear{Heckman et 
al.}{2004}]{2004ApJ...613..109H} Heckman T.~M., Kauffmann G., Brinchmann 
J., Charlot S., Tremonti C., White S.~D.~M., 2004, ApJ, 613, 109 


\bibitem[\protect\citeauthoryear{Hur\'e et 
al.}{1994}]{1994A&A...290...19H} Hur\'e J.-M., Collin-Souffrin S., Le Bourlot J., Pineau des Forets G., 1994, A\&A, 290, 19 

\bibitem[\protect\citeauthoryear{King 
\& Pringle}{2006}]{2006MNRAS.373L..90K} King A.~R., Pringle J.~E., 2006, MNRAS, 373, L90 

\bibitem[\protect\citeauthoryear{King 
\& Pringle}{2007}]{2007MNRAS.377L..25K} King A.~R., Pringle J.~E., 2007, MNRAS, 377, L25 


\bibitem[\protect\citeauthoryear{King, Pringle, 
\& Hofmann}{2008}]{2008MNRAS.385.1621K} King A.~R., Pringle J.~E., Hofmann J.~A., 2008, MNRAS, 385, 1621 


\bibitem[\protect\citeauthoryear{Loska, Czerny, 
\& Szczerba}{2004}]{2004MNRAS.355.1080L} Loska Z., Czerny B., Szczerba R., 2004, MNRAS, 355, 1080 

\bibitem[\protect\citeauthoryear{Moderski, Sikora, 
\& Lasota}{1998}]{1998MNRAS.301..142M} Moderski R., Sikora M., Lasota J.-P., 1998, MNRAS, 301, 142 


\bibitem[\protect\citeauthoryear{Pringle}{1981}]{1981ARA&A..19..137P} Pringle J.~E., 1981, ARA\&A, 19, 137 

\bibitem[\protect\citeauthoryear{Sanders}{1981}]{1981Natur.294..427S} 
Sanders R.~H., 1981, Natur, 294, 427 


\bibitem[\protect\citeauthoryear{Schawinski et 
al.}{2015}]{2015arXiv150506733S} Schawinski K., Koss M., Berney S., Sartori 
L., 2015, arXiv:1505.06733 (SKBS)

\bibitem[\protect\citeauthoryear{Shlosman 
\& Begelman}{1989}]{1989ApJ...341..685S} Shlosman I., Begelman M.~C., 1989, ApJ, 341, 685 

\bibitem[\protect\citeauthoryear{Shlosman, Begelman, 
\& Frank}{1990}]{1990Natur.345..679S} Shlosman I., Begelman M.~C., Frank J., 1990, Nat., 345, 679 





\bibitem[\protect\citeauthoryear{Tombesi et 
al.}{2013}]{2013MNRAS.430.1102T} Tombesi F., Cappi M., Reeves J.~N., Nemmen 
R.~S., Braito V., Gaspari M., Reynolds C.~S., 2013, MNRAS, 430, 1102 

\bibitem[\protect\citeauthoryear{Willott, McLure, 
\& Jarvis}{2003}]{2003ApJ...587L..15W} Willott C.~J., McLure R.~J., Jarvis M.~J., 2003, ApJ, 587, L15 


\end{thebibliography}
\end{document}